\documentclass[aps,prl,twocolumn,showpacs, superscriptaddress, floatfix]{revtex4}
\usepackage[dvipdfmx]{graphicx}

\usepackage{natbib}
\usepackage{amsmath}
\usepackage{amssymb}
\usepackage{bm}

\begin{document}

\title{Directly probing spin dynamics in insulating antiferromagnets using ultrashort terahertz pulses}

\author{P. Bowlan}
\email{pambowlan@lanl.gov}
\affiliation{Center for Integrated Nanotechnologies, MS K771,
Los Alamos National Laboratory, Los Alamos, New Mexico 87545, USA}
\author{S. A. Trugman}
\affiliation{Center for Integrated Nanotechnologies, MS K771,
Los Alamos National Laboratory, Los Alamos, New Mexico 87545, USA}
\author{X. Wang}
\affiliation{Rutgers Center for Emergent Materials and Department of Physics and Astronomy,
Rutgers University, 136 Frelinghuysen Road, Piscataway, New Jersey 08854, USA}
\author{Y. M. Dai}
\affiliation{Center for Integrated Nanotechnologies, MS K771,
Los Alamos National Laboratory, Los Alamos, New Mexico 87545, USA}
\author{S.-W. Cheong}
\affiliation{Rutgers Center for Emergent Materials and Department of Physics and Astronomy,
Rutgers University, 136 Frelinghuysen Road, Piscataway, New Jersey 08854, USA}
\author{E. D. Bauer}
\affiliation{Condensed Matter and Magnet Science Group,  MS K764 Los Alamos National Laboratory, Los Alamos, New Mexico 87545, USA}
\author{A. J. Taylor}
\affiliation{Center for Integrated Nanotechnologies, MS K771,
Los Alamos National Laboratory, Los Alamos, New Mexico 87545, USA}
\author{D. A. Yarotski}
\affiliation{Center for Integrated Nanotechnologies, MS K771,
Los Alamos National Laboratory, Los Alamos, New Mexico 87545, USA}
\author{R. P. Prasankumar}
\email{rpprasan@lanl.gov}
\affiliation{Center for Integrated Nanotechnologies, MS K771,
Los Alamos National Laboratory, Los Alamos, New Mexico 87545, USA}

\date{\today}

\begin{abstract}
We investigate spin dynamics in the antiferromagnetic (AFM) multiferroic TbMnO$_3$ using optical-pump, terahertz (THz)-probe spectroscopy. Photoexcitation results in a broadband THz transmission change, with an onset time of 25~ps at 6~K that becomes faster at higher temperatures. We attribute this time constant to spin-lattice thermalization. The excellent agreement between our measurements and previous ultrafast resonant x-ray diffraction measurements on the same material confirms that our THz pulse directly probes spin order. We suggest that this could be the case in general for insulating AFM materials, if the origin of the static absorption in the THz spectral range is magnetic.	
\end{abstract}

 \maketitle

\section{\label{sec:intro}Introduction}

The ability to switch the magnetization ($M$) in a ferromagnet (FM) on an ultrafast timescale is a longstanding area of fundamental interest, particularly due to its potential applications in magnetic data storage. However, ultrafast control of antiferromagnetic (AFM) materials using femtosecond laser pulses is arguably more promising, since their zero net magnetization makes it easier for the system to change while still conserving the total spin, so that in general their spin dynamics should be much faster than in FMs~\cite{kimel2004laser}. Multiferroic AFM manganites (e.g., RMnO$_3$, where R is a rare earth atom), which can have coexisting and coupled magnetic and ferroelectric orders, have attracted particular interest for their potential device applications~\cite{vopson2015fundamentals}. Improved control and understanding of AFM order in multiferroics could influence practical applications such as four-state memory, ultrafast magnetoelectric switching~\cite{sheu2014using}, or magnetoelectric data storage~\cite{scott2007data}.

Although AFM materials are potentially very useful, their magnetization, especially its temporal evolution, is more difficult to detect, making them less well understood. This is because optical methods for detecting spin order and its dynamics, such as the magneto-optical Kerr effect, are usually only sensitive to a non-zero magnetic moment, $M$~\cite{kimel2004laser}. Optical magnetic linear dichroism can instead be used to probe AFM order~\cite{Bossini2014AFM}; however, the presence of non-magnetic sources of birefringence can make this signal difficult to interpret. Optical second harmonic generation (SHG) is sensitive to AFM spin order and its dynamics~\cite{fiebig2002observation, duong2004ultrafast}, but when applied to multiferroics it must be distinguished from the larger SHG signal originating from ferroelectric order. The other option for probing ultrafast AFM spin dynamics is to use resonant x-ray diffraction with femtosecond x-ray pulses from large scale free electron lasers or synchrotrons~\cite{tobey2012evolution,johnson2015magnetic,kubacka2014large}, which are difficult to gain access to.

Considering that it is still not straightforward to measure ultrafast spin dynamics in AFMs, we recently introduced a simple, table-top method that probes AFM spin dynamics through a magnon resonance using terahertz (THz) pulses~\cite{bowlan2016probing}. Applying this to the AFM multiferroic HoMnO$_3$, after photoexciting electrons with an optical pulse, we observed an induced transparency for the THz probe pulse only at the magnon resonance, clearly indicating a direct sensitivity to spin order. Further analysis of our data showed a change in the magnon line shape (its frequency, amplitude and linewidth) on a timescale of 5-12 picoseconds (ps) that vanished above the Neel temperature $T_{N}$, which was due to spin-lattice thermalization. Furthermore, for temperatures ($T$) less than $T_{N}$, the spin-lattice thermalization time $\tau$ becomes faster with increasing $T$ in HoMnO$_3$, while the opposite happens in the FM manganites, such as La$_{0.7}$Ca$_{0.3}$MnO$_{3}$~\cite{averitt2001ultrafast}. The same trends have also been seen in various types of ultrafast measurements on other AFM manganites~\cite{talbayev2015spin, johnson2015magnetic, qi2012coexistence}. We suggested that this stems from a fundamental difference in FM and AFM systems: lattice vibrations, which conserve the net magnetization, can directly heat spins in an AFM, but not in an FM, which instead requires smaller interaction terms (e.g., spin-orbit coupling) to reduce $M$~\cite{yafet1963}.
	
Here, we further illustrate that THz pulses are a rather general probe of ultrafast spin dynamics, applicable to a broad range of materials with different AFM spin alignments, by applying this technique to a system with a completely different type of spin order. We consider the orthorhombic multiferroic insulator TbMnO$_3$, which is antiferromagnetically ordered below $T_{N1}$ = 42~K, where it is an incommensurate AFM, and also ferroelectric below $T_{N2}$ = 28~K, where it becomes a commensurate AFM~\cite{kenzelmann2005magnetic}. Very similar to that seen in HoMnO$_3$, the photoinduced change in THz transmission has a rise time of $\sim$18-25~ps, which we conclude also comes from spin-lattice thermalization. However, unlike HoMnO$_3$, where the photoinduced changes occurred only at the magnon mode, they happen over the full THz pulse spectrum in TbMnO$_3$. Our data still links this to spin heating, since no Drude response is observed and the broader, flat absorption feature in the THz range is also believed to be magnetic in origin~\cite{takahashi2008evidence, aguilar2007magnon}. Excellent agreement of our results with previous ultrafast resonant x-ray diffraction measurements~\cite{johnson2015magnetic} confirms this. Our results thus further illustrate that THz pulses can provide a direct, table top probe of antiferromagnetic order.

\section{\label{sec:results}Results}
\begin{figure}[tb]
\begin{center}
\includegraphics[width=3.2in]{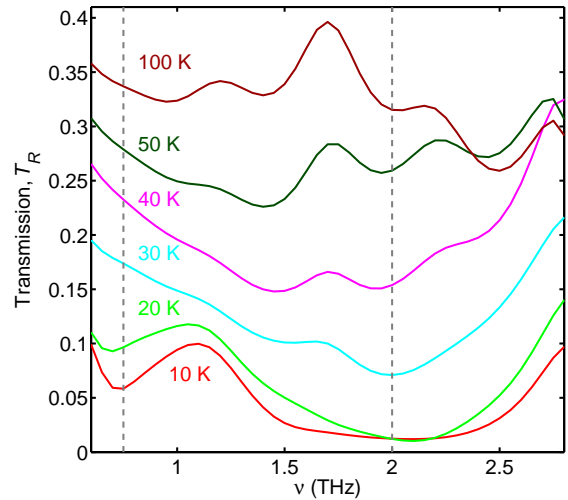}
\caption{ \label{fig:1} (Color online) THz transmission, $T_{R}$, versus frequency $\nu$ (without optical photoexcitation) and sample temperature through a 150~$\mu$m thick $ac$-oriented TbMnO$_3$ crystal, with the THz $E$-field polarized along the crystal $a$-axis. $T_{R}$ is defined as $|E_{trans}(\nu)/E_{in}(\nu)|$, where $E_{in}(\nu)$ is the incident THz pulse and $E_{trans}(\nu)$ is the transmitted pulse. The dotted lines indicate the positions of the electromagnon resonances.}
\end{center}
\end{figure}

The TbMnO$_3$ single crystals used in our experiments were grown in an optical floating zone furnace~\cite{kenzelmann2005magnetic}. The crystal used in our measurements had the $a$ and $c$ axes in-plane and a thickness of $\sim$150~$\mu$m. Our optical-pump, THz-probe (OPTP) system was based on a 1~kHz Ti:sapphire amplifier producing pulses centered at 800~nm, with a duration of $\sim$40 femtoseconds (fs). Ultrashort THz pulses were generated by optical rectification in GaSe and measured by electro-optic sampling in ZnTe (similar to the setup in~\cite{bowlan2012nonlinear}). The THz probe and optical pump pulses were both linearly polarized along the $a$-axis and collinear. The absorption length for the optical pulse is only $\sim$200~nm ~\cite{bastjan2008magneto}, while the THz absorption length is closer to the entire crystal thickness, depending on the frequency and temperature (Fig. \ref{fig:1}); the optical absorption thus determines the effective crystal length for our OPTP measurements. As discussed in our previous work on HoMnO$_3$, lateral diffusion (either heat or transport) is too slow to have any significant effect on our time constants (which might result from the pump and probe penetration depth mismatch)~\cite{bowlan2016probing}. The fact that our THz transmission probe gives a very similar time constant to that measured with x-rays in reflection (discussed in more detail below), where the probe penetration depth is less than that of the pump, confirms this.

Fig.~\ref{fig:1}(a) shows the THz spectra ($T_{R}$) transmitted through our TbMnO$_3$ crystal as a function of temperature ($T$) without optical photoexcitation. No Drude-like response is seen (for comparison see the dashed line in Fig. 2(b)), and instead the THz absorption is dominated by magnetic effects. Below $T_{N2}$, the two well-known electric-dipole active magnon modes in TbMnO$_3$ at $\sim$0.75 and 2~THz are apparent~\cite{pimenov2006possible,aguilar2009origin}. On top of these peaks, there is a continuum-like absorption feature with a full width of about 130~cm$^{-1}$ (~4~THz) which has been attributed to a band of infrared-active, two-magnon excitations~\cite{aguilar2007magnon,takahashi2008evidence,kida2009terahertz}. This broad absorption is seen even above $T_{N1}$, since short range magnetic order has been observed to develop in TbMnO$_{3}$ well above the magnetic ordering temperatures~\cite{takahashi2008evidence,kajimoto2004magnetic}. In Fig.~\ref{fig:1}, the two-magnon absorption band is most apparent for $T>T_{N1}$, manifested as a broad, flat feature which slowly vanishes with increasing temperature (i.e., $T_R$ increases), consistent with \cite{takahashi2008evidence}. The oscillations in the transmission at higher temperatures, where there is less absorption, are from interference between reflections from the surfaces of our $\sim$150~$\mu$m thick crystal.

\begin{figure}[tb]
\begin{center}
\includegraphics[width=3.2in]{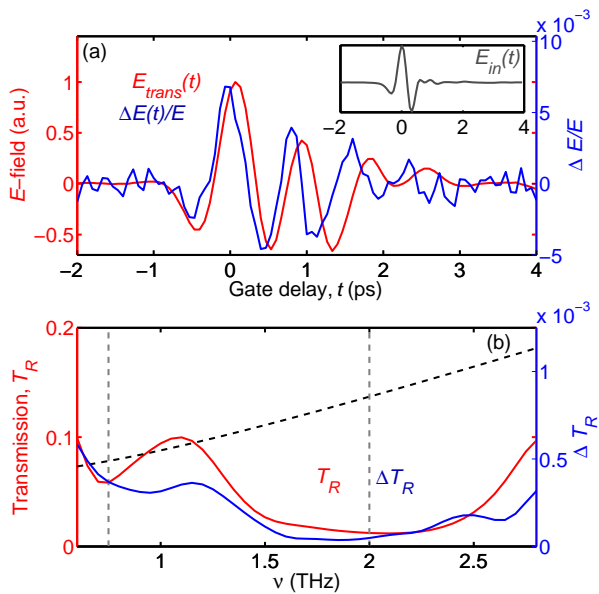}
\caption{\label{fig:2} (Color online) (a)  THz $E$-field versus time transmitted through the crystal (red) ($E_{trans}(t)$) without photoexcitation, and the photoinduced change versus time in the THz $E$-field (blue) ($\Delta E/E = E_{pumped}(t) - E_{trans}(t)$), normalized to the maximum of $E_{trans}$ and measured at $\tau$=100~ps. The inset shows the THz $E$-field before the TbMnO$_3$ crystal, $E_{in}(t)$. (b) The photoinduced THz transmission change (blue), $\Delta T_{R}=|E_{pumped}(\nu) - E_{trans}(\nu)|/|E_{in}(\nu)|$, compared to the steady state transmission (red) (taken from $T_{R}$(10 K) in Fig. \ref{fig:1}). The black dashed line is the calculated transmission of a Drude response for comparison. The vertical, dashed lines indicate the location of the electromagnon resonances. This data was measured at $T\sim$10~K. }
\end{center}
\end{figure}

Fig.~\ref{fig:2}(a) shows the time-dependent electric ($E$) field of the pulse transmitted through the crystal ($E_{trans}(t)$) at 10~K, which exhibits oscillations at later times (as compared to the $E$-field of the incident single cycle THz pulse in the inset of Fig.~\ref{fig:2}(a), $E_{in}(t)$) due to the electromagnon resonances. Next, we optically excited the TbMnO$_3$ crystal at 800~nm with a fluence of $F$=6~mJ/cm$^2$ (corresponding to $\sim10^{22}$ carriers/cm$^3$ at 10~K, or $\sim$0.1 carrier/unit cell). The blue curve in Fig.~\ref{fig:2}(a) shows the resulting photoinduced change in the transmitted THz $E$-field ($\Delta E(t)/E$, defined in the caption of Fig.~\ref{fig:2}) for a pump-probe delay of $\tau$=100~ps.  At early gate delays ($t$ = 0) the photoinduced THz transmission change is in phase with the transmitted THz pulse, representing a spectrally broad increase in transmission. The out-of-phase photoinduced changes at later times, where the oscillations from the electromagnons dominate, indicate that photoexcitation reduces the amplitude of these oscillations, or that the electromagnon absorption is decreasing. Both of these effects are consistent with what would be expected from a steady state temperature increase (see Fig.~\ref{fig:1}). The photoinduced THz transmission change versus frequency, in comparison with the steady state transmission, is shown in Fig.~\ref{fig:2}(b). Also consistent with photoinduced spin heating, no Drude response is seen in the photoinduced transmission change (compare to the black dashed line in Fig.~\ref{fig:2}(b)). The similarity between the static and photoinduced transmission (Fig.~\ref{fig:2}(b)) again shows that the photoinduced changes were spectrally uniform across the THz pulse. Note that at a fluence of 6~mJ/cm$^2$ we estimate $\sim$4 K steady state heating, using the model from ~\cite{demsar2006dynamics} and the thermal conductivity given in ref. ~\cite{berggold2007anomalous}, which is approximately constant over the 10-50 K temperature range that we consider. We have adjusted all of the temperatures in our OPTP data by this amount relative to our cryostat settings.

\begin{figure}[tb]
\begin{center}
\includegraphics[width=3.2in]{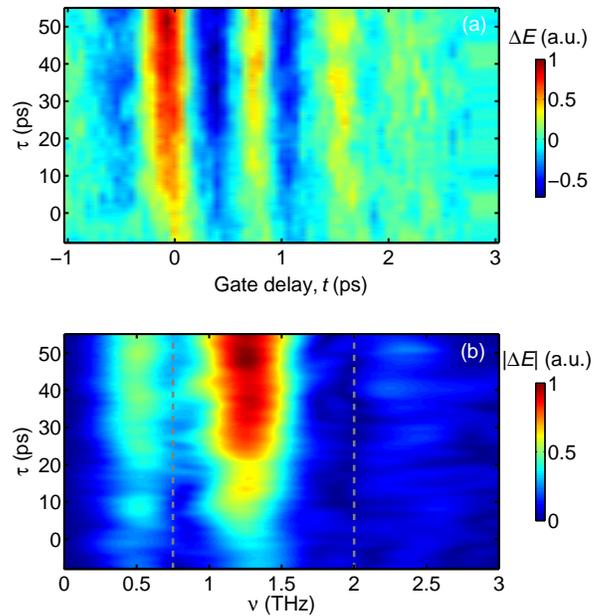}
\caption{ \label{fig:3} (Color online) (a) Photoinduced change in the transmitted THz pulse, $\Delta E$, versus gate delay $t$ and pump-probe delay $\tau$ at $T$=10~K. (b) The Fourier transform of (a), showing $|\Delta E(\nu, \tau)|$, or the power spectrum versus $\tau$.  The grey dashed lines at 0.75 and 2~THz show the positions of the electromagnons.}
\end{center}
\end{figure}

\begin{figure}[tb]
\begin{center}
\includegraphics[width=3.2in]{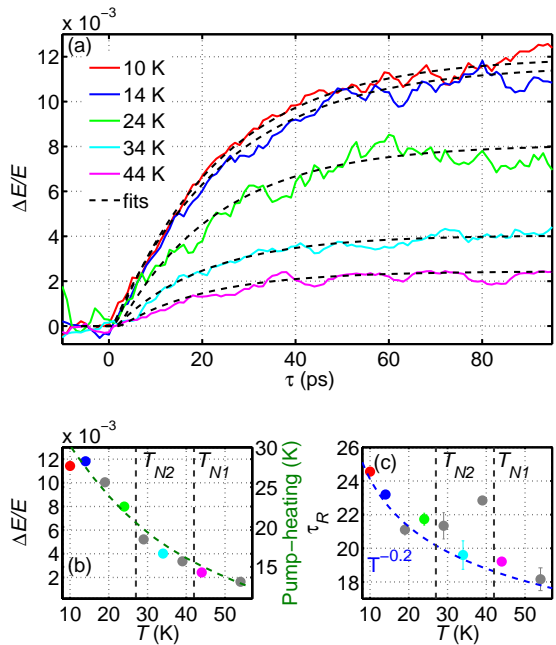}
\caption{ \label{fig:4} (Color online) (a) OPTP signals at different temperatures for $t$=0~ps and $F$ = 6~mJ/cm$^2$. (b) Amplitude of the OPTP signals versus temperature. The right y-axis shows the pump-induced heating (dashed line), calculated from the heat capacity and fluence~\cite{kumar2011specific}. (c) The rise time of the OPTP signals versus $T$, extracted from an exponential fit. The blue dashed line is a power law fit, where the power Q= -0.2. In both (b) and (c), the dashed lines show the transition temperatures $T_{N1}$ and $T_{N2}$.}
\end{center}
\end{figure}

\begin{figure}[tb]
\begin{center}
\includegraphics[width=3.2in]{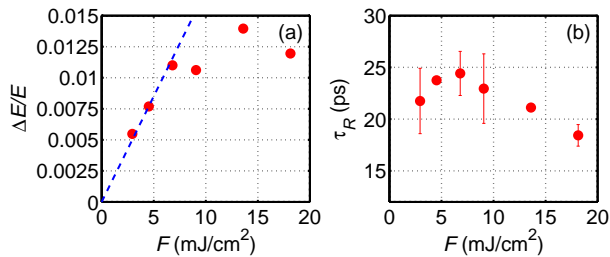}
\caption{ \label{fig:5} (Color online) (a) The amplitude and (b) time constant $\tau_{R}$ extracted from OPTP signals measured at 10~K for different fluences and $t$=1~ps. The dashed line in (a) illustrates the linear dependence of the amplitude on fluence up to 7~mJ/cm$^2$.}
\end{center}
\end{figure}

To investigate the dynamics of the photoinduced transparency, we measured $\Delta E$ versus both gate and pump-probe delays at $T$=10~K, as shown in Fig.~\ref{fig:3}(a). The power spectrum of Fig.~\ref{fig:3}(a) is shown in Fig.~\ref{fig:3}(b). These images show that for all gate delays, or all frequencies across the transmitted THz pulse, the photoinduced transparency has the same pump-probe delay dependence and builds up over $\sim$25~ps. No further changes were seen up to the latest time delays of $\sim$300~ps. Fig.~\ref{fig:4}(a) shows pump-probe signals for different sample temperatures versus delay $\tau$, measured at a fixed gate delay where the difference was largest ($t$~=~0~ps in Fig.~\ref{fig:2}). The amplitude ($\Delta E/E$) of the signals at $\tau$ = 100~ps for different temperatures is shown in Fig.~\ref{fig:4}(b). The right $y$-axis (dashed line) of this figure shows the instantaneous pump-induced heating, calculated from the heat capacity~\cite{kumar2011specific} and the pump fluence. The shape of this curve is in good agreement with the amplitude of the pump-probe signal, strongly suggesting that the phonon temperature determines the amount of spin heating. In addition, using a single exponential fit, we extracted the temperature-dependent rise time of the curves in Fig.~\ref{fig:4}(a), shown in Fig.~\ref{fig:4}(c). We fit the temperature dependence of the resulting time constants to a power law, $T^Q$, finding $Q=-0.2$. Finally, the fluence dependence of the time constant $\tau_R$ and the amplitude of the OPTP signals are plotted in Fig.~\ref{fig:5} for $T$=10~K, $t$=0~ps.

\section{\label{sec:discussion}Discussion}

Several features of our data indicate that THz pulses can directly probe spin order in TbMnO$_3$. In our previous work on HoMnO$_3$~\cite{bowlan2016probing} this was more obvious since the photoinduced changes occurred only at the magnon resonance, while in TbMnO$_3$, the spectral changes happen across the whole THz spectrum (Fig.~\ref{fig:2}(b)). Here, the most compelling evidence comes from a comparison of our measurements to a recent optical-pump, resonant x-ray diffraction study on TbMnO$_3$~\cite{johnson2015magnetic}, since that method is already known to be a direct, ultrafast probe of spins~\cite{tobey2012evolution}. Our THz data agrees very well with those measurements. In both cases a single exponential rise was observed (Fig.~\ref{fig:4}) with a time constant of $\sim$23~ps for $T$=11-12~K and $F$= 6~mJ/cm$^2$. The x-ray study indicated that this was associated with a melting of spin order (through heating the spin system). Our measured fluence dependence, shown in Fig.~\ref{fig:5}, also agrees well with the trends shown in the resonant x-ray diffraction study (see Fig.~3 in ref. \cite{johnson2015magnetic}), since in both cases the amplitude of the OPTP signal saturates at $\sim$6 mJ/cm$^2$. We note that the time constant drops more rapidly with increasing fluence in the x-ray data~\cite{johnson2015magnetic}, since this approach was only sensitive to long-range spin order, while our THz probe is also sensitive to short-range spin order (relevant at higher temperatures and fluences).
	
Further evidence for the fact that we probe spin dynamics across our entire THz spectral window comes from the continuum-like infrared active two-magnon excitation in the static THz absorption of TbMnO$_3$ discussed above~\cite{takahashi2008evidence}. This indicates that the observed static absorption features, including the electromagnons and the broad, flat background seen in Fig.~\ref{fig:1}, are all magnetic in origin (and hence the dynamics are too). This is also supported by the fact that the OPTP signal in TbMnO$_3$ persists above $T_{N1}$ and slowly decreases above this temperature (it was no longer detectable above 54~K) ~\cite{takahashi2008evidence,aguilar2007magnon}. Finally, if TbMnO$_3$ were not a good insulator, photoexcited electron-hole pairs would likely lead to a Drude response in the THz conductivity spectrum, which might dominate or obscure the spin dynamics. As shown in Fig.~\ref{fig:2}(c), a photoinduced Drude response was not observed. Therefore, we suggest that in general, THz pulses may be able to probe ultrafast spin dynamics in any insulating AFM system, as long as the static THz absorption is of magnetic origin.

Next, we argue that the time constant $\tau_R$ observed in our OPTP measurements is due to spin-lattice thermalization. Our 800~nm pump pulse photoexcites intersite Mn-Mn electron transitions in orthorhombic TbMnO$_3$. Given that we observe spin heating in our OPTP measurements, and that the energy from optical photoexcitation is deposited in the electrons, the next question is how electrons transfer energy to the spins. Our THz data in Figs.~\ref{fig:3} and \ref{fig:4} does not show the initial fast transfer of energy from electrons to phonons that is normally observed in most materials, most likely because of our limited time resolution ($\sim$250~fs). However, optical-pump/optical-probe measurements on TbMnO$_3$, which are directly sensitive to electronic order, show that this process occurs within $\sim$30-100~fs of photoexcitation ~\cite{qi2012coexistence,handayani2013dynamics}; this is typical for manganites (see, e.g., ~\cite{wang2013understanding, wall2009ultrafast}). Therefore, before the $\sim$18-25~ps relaxation process described by $\tau_R$ occurs, the electrons and phonons have already thermalized, pointing towards a phonon-mediated transfer of energy from electrons to spins (also often seen in manganites~\cite{averitt2002ultrafast}). Further evidence for this comes from the ultrafast lattice heating calculations in Fig.~\ref{fig:4}(b), showing that our estimate of the photoinduced lattice temperature increase agrees well with the measured spin temperature increase shown in Fig. 4(a) of ref.~\cite{johnson2015magnetic}, both of which are $\sim$~27~K. The above considerations thus indicate that after the relaxation process shown in Figs.~\ref{fig:4}(c) and \ref{fig:5}(b), the spins and lattice are in thermal equilibrium, allowing us to ascribe $\tau_R$ to spin-lattice thermalization (as in ref. ~\cite{bowlan2016probing}). We suggest that the importance of short-range magnetic order in TbMnO$_3$~\cite{kajimoto2004magnetic,takahashi2008evidence}, apparent also from the fact that our OPTP signal persists above $T_{N1}$, as well as the temperature dependence of the broad two-magnon excitation, could account for the slower spin-lattice relaxation in TbMnO$_3$ as compared to HoMnO$_3$~\cite{bowlan2016probing}, where such a feature was not present. This is also consistent with all-optical pump-probe measurements on Eu$_{0.75}$Y$_{0.35}$MnO$_{3}$, a system also known to have short range magnetic order above $T_{N}$~\cite{aguilar2007magnon}, which showed a relaxation time similar to that seen here that also slowly decreases in amplitude above $T_{N}$~\cite{talbayev2015spin}.

We now discuss the microscopic mechanism by which spin-lattice thermalization could occur in TbMnO$_3$ and other AFM systems. Previous ultrafast studies on orthorhombic AFM RMnO$_3$ compounds~\cite{talbayev2015spin, johnson2015magnetic} proposed that the formation of optically induced polarons could play a role in ultrafast spin heating. As described above, in the orthorhombic manganites, absorption of light at 800~nm is associated with Mn-Mn inter-site transitions, changing the charge of the ions and resulting in the formation of small polarons~\cite{allen1999anti}. In contrast, optical excitation of hexagonal manganites such as HoMnO$_3$ results in on-site transitions and a very small perturbation of the polaronic potential, yet a similar spin-lattice relaxation time was observed in this system~\cite{bowlan2016probing}. Also, the formation and relaxation of optically excited polarons observed in other manganites and semiconductors typically takes place in $<$~1~ps~\cite{prasankumar2007phase, gaal2007internal,wu2009ultrafast} due to the small spatial scales involved. Considering these facts and that a range of different AFM systems all show similar monotonically decreasing spin-lattice thermalization times with temperature~\cite{bowlan2016probing,qi2012coexistence,chia2006quasiparticle,talbayev2015spin,handayani2013dynamics}, we suggest that instead of involving excitations associated with a specific type of spin or lattice order (such as polarons), the microscopic mechanism governing spin-lattice relaxation in these compounds could instead more generally be related to the fact that they are all AFMs.

To learn more about this trend in AFMs, we consider the commonly used two-temperature model (TTM) for the spin-lattice thermalization time, $\tau_{SL}$. Ref.~\cite{groeneveld1995femtosecond} shows that this is given by $C_{s}/g$, where $C_s$ is the spin specific heat and $g$ is the spin-lattice coupling constant, under the assumption that $C_s$ is much smaller than the lattice specific heat. In FMs, one can assume that $g$ has no temperature dependence~\cite{beaurepaire1996ultrafast}, since $\tau_{SL}(T)$ follows the temperature dependence of $C_{s}(T)$~\cite{lobad2000spin,averitt2001ultrafast}, and therefore, like $C_{s},$ peaks at the Curie temperature. Similarly, $\tau_R$ in TbMnO$_3$ (Fig.~\ref{fig:4}(c)), as well as some of the other AFM systems discussed above~\cite{qi2012coexistence,talbayev2015spin}, also shows peaks at the Ne\'el temperatures $T_{N1}$ and $T_{N2}$, following the peaks in $C_s$~\cite{kumar2011specific}. However, a strong monotonic decrease in $\tau_R$ with temperature is also seen on top of this, as described by the power law fit shown in Fig. \ref{fig:4}(c). This suggests that, unlike the FM manganites, the spin-lattice coupling constant $g$ has a stronger, non-negligible temperature dependence in AFMs, pointing to a fundamental difference in the way that spins and phonons couple in these two types of systems.

As discussed in our previous paper~\cite{bowlan2016probing}, we can gain insight into differences in spin-lattice thermalization between FM and AFM systems by considering that the simplest Hamiltonian for spin-lattice thermalization involves an exchange constant $J(r)$, where $r$ is the distance between atoms. Through this type of interaction, phonons exchange energy with spins by directly modulating $J(r)$. This Hamiltonian conserves the net magnetization $M$ and can therefore heat spins in AFMs, where $M=0$, even in the spin ordered state. However this Hamiltonian cannot account for spin heating in FMs, in which $M\neq$ 0; spin-lattice relaxation in this case happens instead through interactions that reduce $M$, such as spin-orbit coupling~\cite{yafet1963}. Consistent with this argument, spin-orbit coupling is usually weak in RMnO$_3$ compounds because of crystal field quenching, making spin-lattice thermalization slower in the FM manganites~(e.g.,~\cite{lobad2000spin,averitt2001ultrafast}). Therefore, another possible, potentially larger microscopic mechanism for spin-lattice thermalization in AFMs (in addition to spin-orbit coupling) is direct heating of spins by phonons through the exchange interaction.

To test this idea, we follow the Boltzmann rate equation model of~\cite{abrahams1952spin}, based on a Hamiltonian for spin-lattice thermalization with direct coupling through the exchange interaction $J(r)$. Although this was originally intended for FMs, before it was shown that a more complex Hamiltonian was needed~\cite{yafet1963}, we propose that it could apply to AFMs, where this term may dominate over spin-orbit coupling. Specifically, this interaction is computed by considering the leading magnon-phonon scattering process, which in this case, is one phonon creating one magnon and annihilating another. The calculation therefore depends on the magnon and phonon dispersions of the material, as well as the populations of these particles, which gives the resulting thermalization time a strong temperature dependence. Applying this model to HoMnO$_3$, we were able to reproduce a monotonically decreasing $\tau_{SL}$ for $T<T_{N}$, as in our measurements (both here and in ~\cite{bowlan2016probing}). However, the calculated exponential describing the temperature dependence was $T^{-3}$, which is faster than the measured $T^{-0.5}$ dependence in HoMnO$_3$; because of this discrepancy we have not yet attempted to apply this to TbMnO$_3$, but would expect a similar outcome. Some possible reasons for the discrepancy are that the model assumes that the spin and phonon subsystems thermalize instantaneously amongst themselves after scattering and that the magnetic anisotropy is temperature independent, which neutron studies on other manganites suggest may not be the case~\cite{vajk2005magnetic}. It may also be important to include spin-orbit coupling and consider the insulating nature of the AFM manganites, unlike most FMs. In the future, we plan to develop a more detailed quantitative model for spin-lattice relaxation in AFM manganites based on these ideas.

\section{\label{sec:summary}Summary}

We demonstrated that in insulating AFMs, THz pulses can directly probe spin order. Applying this to the AFM multiferroic TbMnO$_3$, we observed an optically induced transmission change that developed within 18-25~ps after photoexcitation. Excellent agreement with a previous ultrafast x-ray study confirms that we directly probe spin order, and that the observed dynamics originate from spin-lattice thermalization, as in our previous study of HoMnO$_3$~\cite{bowlan2016probing}. However, the current study enables us to go beyond this and point out that many different AFM systems show very similar spin-lattice thermalization times with similar temperature dependencies, where $\tau_{SL}$ decreases with a power law dependence on temperature, but like FMs still shows peaks at the transition temperatures. We explain this in terms of a strongly temperature-dependent spin-lattice coupling constant for AFMs (unlike in FMs), which could stem from the fact that lattice vibrations can directly heat spins in AFMs. This work thus demonstrates a powerful approach for directly probing AFM spin dynamics in insulators, applicable to a wide range of systems, and gives new clarity to previous ultrafast measurements made on such systems. More generally, the idea of probing the ultrafast dynamics of order parameters through low energy resonances is applicable to phonons as well as magnons and can shed new light on the couplings between these resonances, which will be especially useful in unraveling the physics of correlated electron systems.

\begin{acknowledgments}
The ultrafast measurements were performed at the Center for Integrated Nanotechnologies, a U.S. Department of Energy, Office of Basic Energy Sciences user facility and also partially supported by the NNSA's Laboratory Directed Research and Development Program. Los Alamos National Laboratory, an affirmative action equal opportunity employer, is operated by Los Alamos National Security, LLC, for the National Nuclear Security Administration of the U. S. Department of Energy under contract DE-AC52-06NA25396. The work performed at Rutgers University was supported by the DOE under Grant No. DOE: DE-FG02-07ER46382.  We thank Rolando Vald\'es Aguilar for helpful discussions.
\end{acknowledgments}

\providecommand{\noopsort}[1]{}\providecommand{\singleletter}[1]{#1}%

\end{document}